\documentstyle[graphicx]{mn}

\newif\ifAMStwofonts

\ifoldfss
  \newcommand{\rmn}[1] {{\rm #1}}

  \ifCUPmtlplainloaded \else
    \NewTextAlphabet{textbfit} {cmbxti10} {}
    \NewTextAlphabet{textbfss} {cmssbx10} {}
    \NewMathAlphabet{mathbfit} {cmbxti10} {} 
    \NewMathAlphabet{mathbfss} {cmssbx10} {} 
  \fi
  \ifAMStwofonts
    \ifCUPmtlplainloaded \else
      \NewSymbolFont{upmath} {eurm10}
      \NewSymbolFont{AMSa} {msam10}
      \NewMathSymbol{\upi}     {0}{upmath}{19}
      \NewMathSymbol{\umu}     {0}{upmath}{16}
      \NewMathSymbol{\upartial}{0}{upmath}{40}
      \NewMathSymbol{\leqslant}{3}{AMSa}{36}
      \NewMathSymbol{\geqslant}{3}{AMSa}{3E}

    \fi
  \fi
\fi 

\ifnfssone
  \newmathalphabet{\mathit}
  \addtoversion{normal}{\mathit}{cmr}{m}{it}
  \addtoversion{bold}{\mathit}{cmr}{bx}{it}
  \newcommand{\rmn}[1] {\mathrm{#1}}

  \newmathalphabet{\mathbfit} 
  \addtoversion{normal}{\mathbfit}{cmr}{bx}{it}
  \addtoversion{bold}{\mathbfit}{cmr}{bx}{it}
  \newmathalphabet{\mathbfss} 
  \addtoversion{normal}{\mathbfss}{cmss}{bx}{n}
  \addtoversion{bold}{\mathbfss}{cmss}{bx}{n}
  \ifAMStwofonts
    \ifCUPmtlplainloaded \else
      \UseAMStwoboldmath
      \makeatletter
      \new@mathgroup\upmath@group
      \define@mathgroup\mv@normal\upmath@group{eur}{m}{n}
      \define@mathgroup\mv@bold\upmath@group{eur}{b}{n}
      \edef\UPM{\hexnumber\upmath@group}
      \new@mathgroup\amsa@group
      \define@mathgroup\mv@normal\amsa@group{msa}{m}{n}
      \define@mathgroup\mv@bold\amsa@group{msa}{m}{n}
      \edef\AMSa{\hexnumber\amsa@group}
      \makeatother
      \mathchardef\upi="0\UPM19
      \mathchardef\umu="0\UPM16
      \mathchardef\upartial="0\UPM40
      \mathchardef\leqslant="3\AMSa36
      \mathchardef\geqslant="3\AMSa3E
    \fi
  \fi
\fi 

\ifnfsstwo
  \newcommand{\rmn}[1] {\mathrm{#1}}

  \DeclareMathAlphabet{\mathbfit}{OT1}{cmr}{bx}{it}
  \SetMathAlphabet\mathbfit{bold}{OT1}{cmr}{bx}{it}
  \DeclareMathAlphabet{\mathbfss}{OT1}{cmss}{bx}{n}
  \SetMathAlphabet\mathbfss{bold}{OT1}{cmss}{bx}{n}
  \ifAMStwofonts
    \ifCUPmtlplainloaded \else
      \DeclareSymbolFont{UPM}{U}{eur}{m}{n}
      \SetSymbolFont{UPM}{bold}{U}{eur}{b}{n}
      \DeclareSymbolFont{AMSa}{U}{msa}{m}{n}
      \DeclareMathSymbol{\upi}{0}{UPM}{"19}
      \DeclareMathSymbol{\umu}{0}{UPM}{"16}
      \DeclareMathSymbol{\upartial}{0}{UPM}{"40}
      \DeclareMathSymbol{\leqslant}{3}{AMSa}{"36}
      \DeclareMathSymbol{\geqslant}{3}{AMSa}{"3E}
    \fi
  \fi
\fi 

\ifCUPmtlplainloaded \else
  \ifAMStwofonts \else 
    \def\upi{\pi}
    \def\umu{\mu}
    \def\upartial{\partial}
  \fi
\fi

\title[Submillimetre continuum observations of Cygnus-A]{Millimetre and submillimetre continuum observations of the 
core and hotspots of Cygnus-A}

\author[E.\ I.\ Robson et al.]
       {E.\ I.\ Robson,$^1$ L.\ L.\ Leeuw,$^{1,2}$ J.\ A.\ Stevens$^3$ and
       W.\ S.\ Holland$^1$ 
\\ $^1$ Joint Astronomy Centre, 660 N. A'oh\=ok\=u Place,
       University Park, Hilo, HI 96720, USA  
\\ $^2$ Centre for Astrophysics, University of Central Lancashire,
Preston PR1 2HE
\\ $^3$  Mullard Space Science Laboratory, University College
London, Holmbury St. Mary, Dorking, Surrey RH5 6NT}
\date{Accepted 1998 July 29.
      Received 1998 July 20; in original form 1998 June 8}

\pagerange{\pageref{firstpage}--\pageref{lastpage}}
\pubyear{1998}

\begin{document}

\maketitle

\label{firstpage}

\begin{abstract}
We present millimetre photometry and submillimetre imaging of 
the central core and two hotspots in the radio lobes of the galaxy 
Cygnus-A. For both hotspots and the central core the 
synchrotron spectrum continues smoothly from the radio to a 
frequency of 677\,GHz.  The spectral index of the hotspots is constant over our frequency range,
with a spectral index of $\alpha
\approx -1.0$ (S$_{\nu} \propto \nu^{\alpha} $), which is steeper than at lower
frequencies and represents the emission from an aged population
of electrons. The core is significantly flatter, with $\alpha = -0.6 \pm 0.1$,
suggestive of an injected spectrum with no ageing, but some evidence for
steepening exists at our highest observing frequency.  Although $IRAS$ data suggest the presence of dust in Cygnus-A, 
our 450\,$\mu{\rm m}$ data show no evidence of cold dust, therefore 
the dust component must have a temperature lying between 85\,K 
and 37\,K, corresponding to dust masses of $1.4 \times 10^6 
\rm{M}_{\odot}$ and $1.0 \times 10^8$\,{\rm M}$_{\odot}$ respectively. 
\end{abstract}


\section{Introduction}

Cygnus-A is locally the most powerful Fanaroff-Riley (FRII) radio 
galaxy and therefore it is also the best studied in terms of 
spatial resolution. It has even had an entire workshop 
devoted to it and global properties of the object are well 
reviewed in these references (Carilli \& Barthel 1996, Carilli \&
Harris 1996, and Carilli et al.\, 1998).  Cygnus-A has featured 
prominently in the quest for the unification of powerful 
radio galaxies and quasars, being cast as the classical case of a 
quasar in the plane of the sky (Antonnuci \& Miller 1985, Barthel 1989). In terms of the 
synchrotron emission, at low radio frequencies the two giant 
lobes dominate the emission (Hargrave \& Ryle 1974), but at higher frequencies, 
the hotspots, or working surfaces in the lobes, become more 
prominent along with the galaxy core.  The southern and northern hotspots are respectively $50''$ and $70''$ from the core.  Using a 
Hubble constant of 75\,km\,s$^{-1}$ Mpc$^{-1}$ and a redshift of $z = 0.0562$ 
(Stockton et al. 1994), Cygnus-A lies at a distance of 227\,Mpc.  

Deep VLA images reveal the thin jet which transports energy 
from the AGN core to the radio lobes (Perley, Dreher, Cowan 1984).
VLBI observations by Linfield (1984) and other authors, and most
recently by Krichbaum et al.\
(1998), have extended this picture at sub-milliarcsecond resolution.  The fact that the 
electron synchrotron lifetime in the hotspots is less than the 
light travel time from the central core (Hargrave \& Ryle 1974) means that electron re-
acceleration must take place in the lobes, and the hotspots are 
believed to be the working surfaces at which this re-
acceleration takes place. Indeed, the prominence of hotspots 
at the outer edges of the lobes is one of the identifying 
features of FRII radio sources (Fanaroff \& Riley 1974). The precise mechanism 
for the electron acceleration is uncertain and determination of 
the synchrotron spectral index is important as 
it can rule out some potential mechanisms. 

For a steady injection model (Bell 1978) a steepening of 0.5 in the 
index is expected at a certain frequency as the electrons lose 
energy more rapidly than is being supplied by the acceleration 
process. At some higher frequency, the electron spectrum will 
cut-off as the high energy electrons rapidly lose energy and 
become depleted; this produces the final turnover and 
steep downturn of the synchrotron emission. The emission 
from the hot-spots (denoted by A and D in the convention 
derived from VLA maps, where A is the north-west and D is 
the south-east hot-spot) has been well observed at a number 
of radio frequencies (Muxlow et al.\
1988, Wright and Birkinshaw 1984, Carilli et
al\ 1991, Salter et al.\ 1989, Eales, Alexander \& Duncan 1989).  Both
are found to  
have an excellent power-law spectral index with a coefficient 
of about $-0.5$ between 0.1\,GHz to around a few GHz, 
steepening to an index of $-1.0$ at higher frequencies. 
The break to the higher index occurs at around 10\,GHz for both hotspots. The spectral break 
of 0.5 is indicative of the steady injection model with constant 
radiative losses.

The spectral shape of the core of Cygnus-A is much less well 
determined than the hotspots. The general shape is of a 
power-law, which is self-absorbed below a few GHz,  and 
above which it has an index of around -0.1 but with significant 
uncertainty (Salter et al.\ 1989, Eales et al.\ 1989).  

In this paper we present high frequency (150 GHz - 857 GHz) 
photometry of the hotspots A and D (we see no evidence for a 
separate component B at our spatial resolution) and the 
galaxy core.  Single-pixel photometry data are given for frequencies 
of 150 GHz, and 222 GHz, and we present the first submillimetre imaging photometry at frequencies of 353 GHz, 400 GHz, 667 GHz and 
857 GHz (corresponding to wavelengths of 2000, 1350, 850,  
750, 450, and 350\,$\mu{\rm m}$, respectively).  

\section[]{Observations}

The imaging and photometry observations were made with SCUBA, the
Submillimetre Common-User Bolometer Array (Holland et al.\, 1998,
Robson et al.\, in preparation) at the
Nasmyth focus of the James Clerk Maxwell Telescope on 1997 May 02, Sep
18, Oct 02 and 1998 Feb 14 and 16.  SCUBA has a 91 element short-wave
(SW) array that is optimized to operate at 450\,$\mu{\rm m}$ or
350\,$\mu{\rm m}$ 
and a 37 element long-wave (LW) array that is optimized to operate at
850\,$\mu{\rm m}$ or 750\,$\mu{\rm m}$.  Both arrays have a 2.3 arcmin instantaneous
field-of-view, and the optics are designed such that with a suitable
jiggle pattern of the secondary mirror of the telescope, fully sampled
maps can be obtained simultaneously at 450 and 850\,$\mu{\rm m}$ or
350 and 750\,$\mu{\rm m}$.  Using specific single bolometers (eg. the central
ones) of the LW and SW arrays and a simpler jiggle pattern, faster
photometric mode observations can instead be obtained simultaneously
at 450 and 850\,$\mu{\rm m}$ or 350 and 750\,$\mu{\rm m}$.
Observation were made in this mode only on 1998 Feb 14 in order to go
deeper and reduce our errors, especially at 450\,$\mu{\rm m}$.  SCUBA also has single
element bolometers optimized to operate at 1100\,$\mu{\rm m}$,
1350\,$\mu{\rm m}$ 
and 2000\,$\mu{\rm m}$. All 131 bolometers are cooled to less than 0.1\,K making
SCUBA limited by the sky background radiation.

For the imaging observations of Cygnus A, the telescope was 
pointed at the core of the optical galaxy using radio positions 
from Hargrave \& Ryle (1974). The radio 
hotspots are about 65\,arcseconds from the core, thereby nicely 
fitting within the instantaneous SCUBA field-of-view. The 
imaging observations employed a 64-point jiggle pattern with a 3\,arcsecond
offset between each position, thereby giving fully  
sampled images with both arrays over a period of 128\,s (1 
integration).  Each integration was divided into 4 so that the 
telescope could be nodded between the signal and reference 
beam every 16\,s. The secondary was chopped in azimuth at 7.8\,Hz and 
with a throw of 120\,arcseconds.

For photometry mode observation the telescope was pointed at 
the radio core position, and at each hotspot using the positions 
from Salter et al.\ (1989). Typically the telescope was slewed to the 
core position and a pointing observation taken to ensure 
excellent positioning; photometry observations were then made on the core 
and the hotspots. The
photometry observations employed a 9-point jiggle pattern in a 3 
by 3 grid of 2\,arcsecond spacing taking 9\,s on the signal and 9\,s
on the
reference beam. In this 
case the secondary had a chop throw of 90\,arcseconds in azimuth.

For all observations the focus stability was checked every three
hours, or more often when there were significant  ambient temperature changes. 
Skydips with SCUBA were performed regularly to measure the 
atmospheric opacity at the two array wavelengths, while the 225\,GHz
opacity was monitored continuously from the Caltech  
Submillimeter Observatory sky monitor (commonly known as 
CSO taumeter, see Masson 1993) to dictate the frequency of 
undertaking SCUBA skydips. Flux calibration was taken from 
planets and the JCMT secondary calibrator list. 

\begin{figure*}
\begin{centering}
\resizebox{4.0in}{!}{\rotatebox{270}{\includegraphics[2.7cm,3.7cm][16cm,23cm]{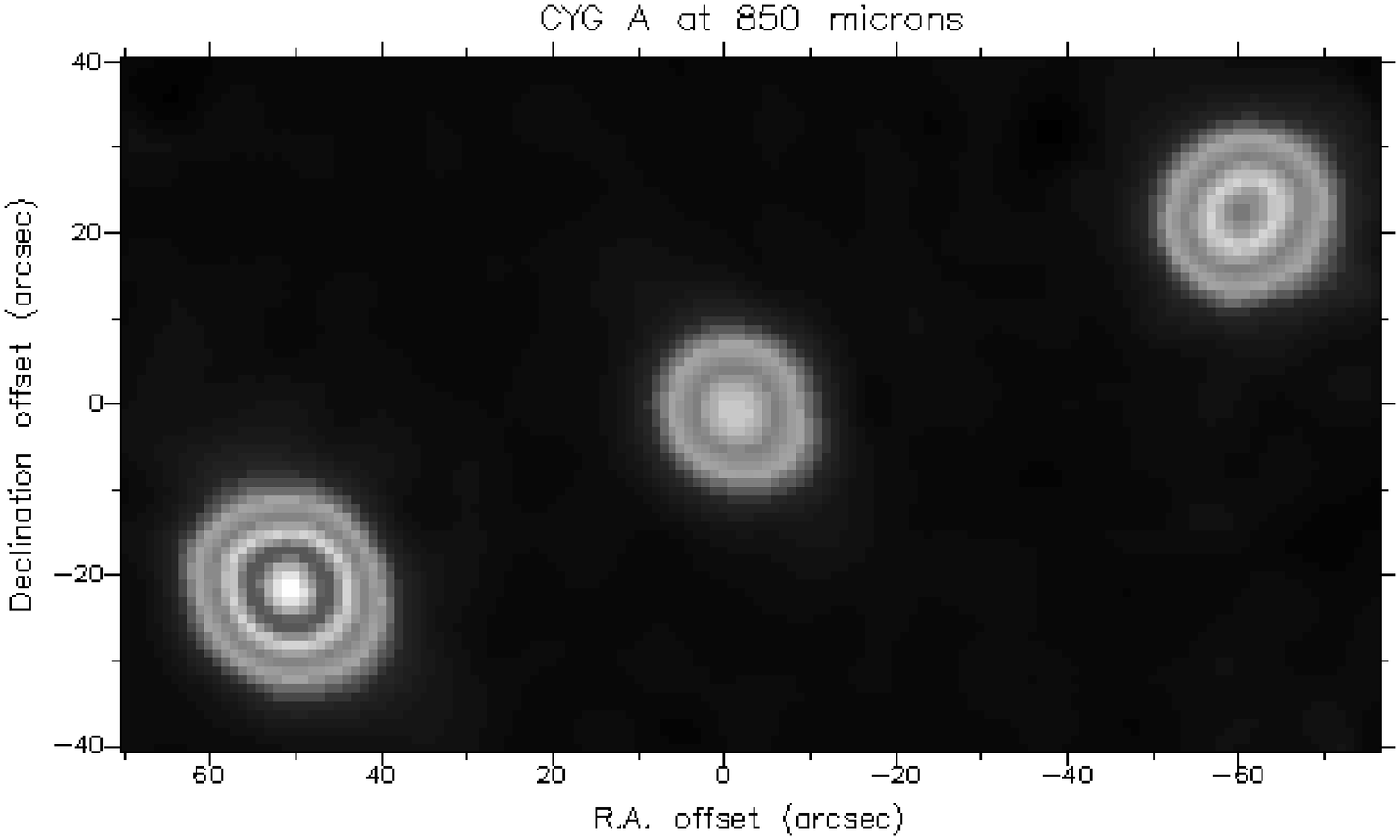}}}
\resizebox{4.0in}{!}{\rotatebox{270}{\includegraphics[3cm,4cm][16cm,23cm]{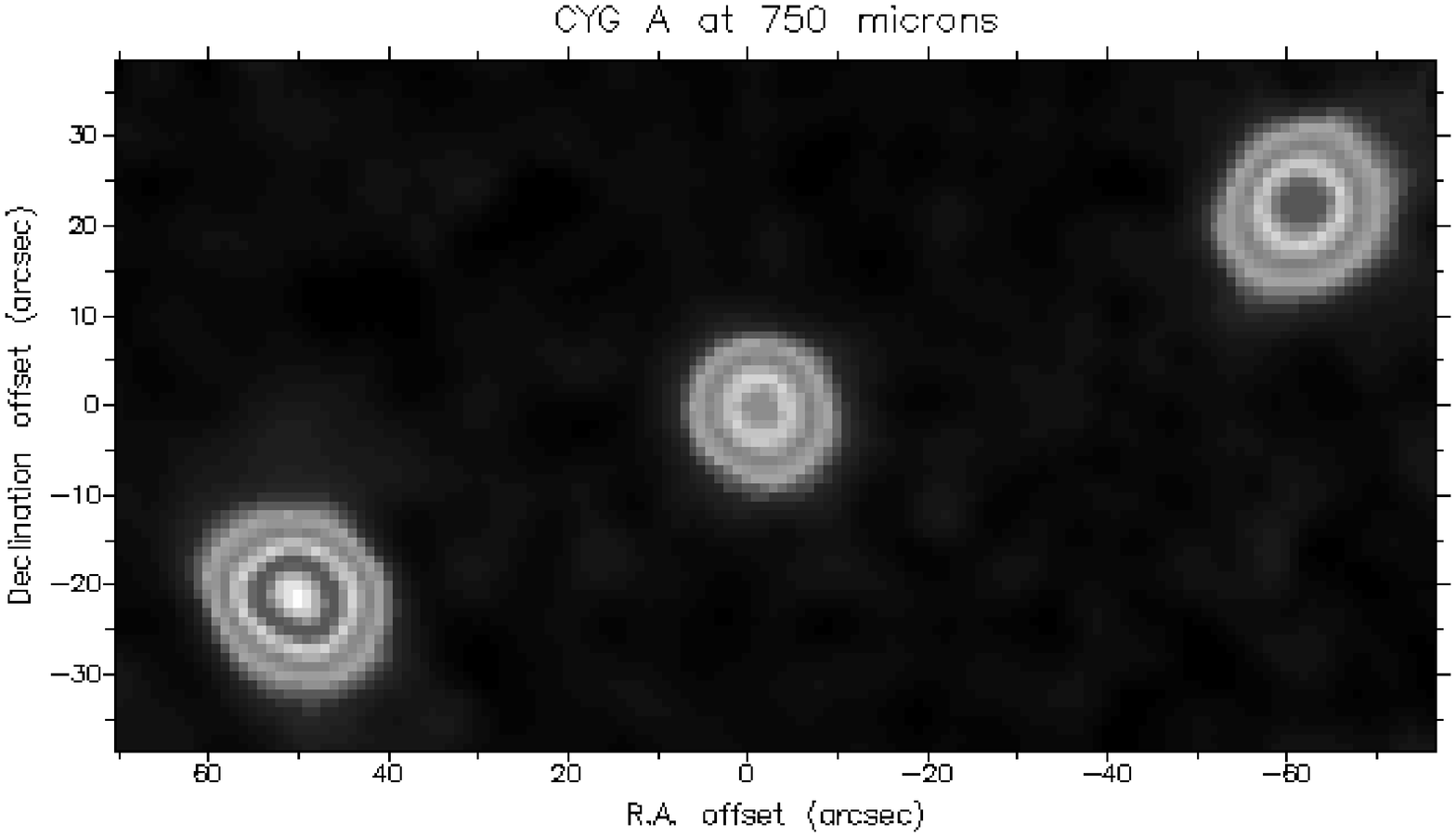}}}
\resizebox{4.0in}{!}{\rotatebox{270}{\includegraphics[3cm,4cm][16cm,23cm]{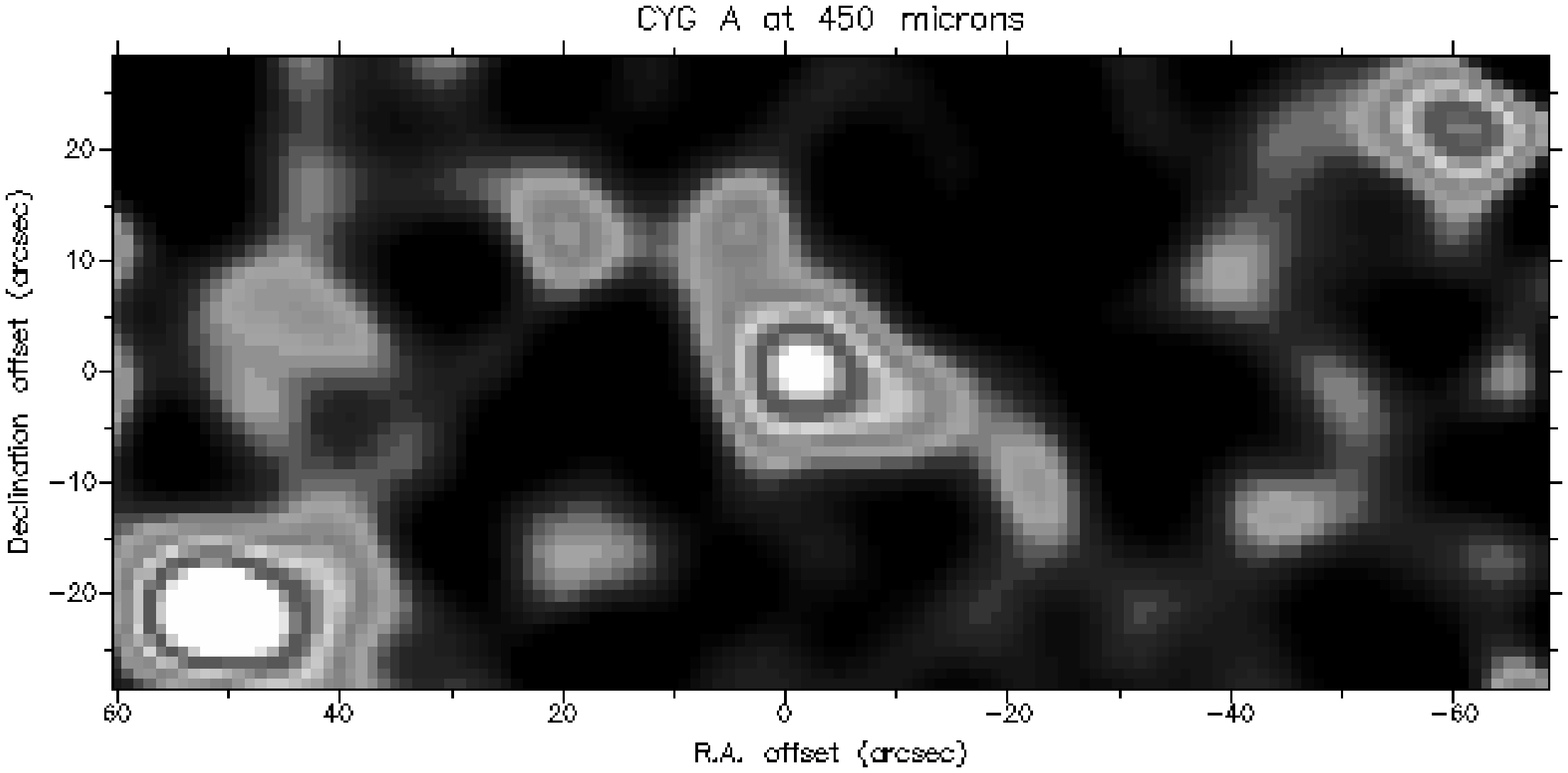}}}
\label{bgfig}
\end{centering}
\caption{The hotspots and central core at 850, 750 and 450\,$\mu$m}
\end{figure*}

The imaging data analysis was undertaken using the dedicated 
SCUBA data reduction software SURF (Jenness et al.\ 1998).  The 
data reduction consisted of first flatfielding the array images, and 
then correcting for atmospheric extinction.  Next, sky-noise 
removal was performed using array bolometers in which there was no source emission. This was 
usually the bolometers in the first ring for the LW array and the 
second ring for the SW array. Noisy bolometers and integrations (or 
subdivisions thereof) were blanked and the images were 
despiked. The resulting data were calibrated using instrumental 
gains that were determined from beam maps of Mars or Uranus 
nightly and in the same observation mode as the target 
observation. During the 1998 Feb runs, the planets were not 
available and the JCMT secondary calibrators CRL 618, CRL 
2688 and IRC+10216 were used
(Sandell 1994; Sandell et al.\ in preparation).  

For the 1350\,$\mu{\rm m}$ and 2000\,$\mu{\rm m}$ photometry 
observations no sky-noise removal was done, as the observation used 
only a single bolometer, and spike removal was performed by 
clipping the data at a specified sigma. Flux calibration was made 
by photometry observations of Mars and Uranus.

The results from the imaging and photometry observations are 
shown in Table 1. The uncertainties are a quadratic sum of 
the uncertainty arising from the measured signal-to-noise-ratio and a
systematic calibration uncertainty, which varies from 10\% at
850\,$\mu{\rm m}$ and lower, to 27\% at  450\,$\mu{\rm m}$. 


\begin{table*}
 \centering
 \begin{minipage}{140mm}
  \caption{SCUBA Observations}
\begin{tabular}{lrrrrrrrrrr}
 & \multicolumn{10}{c}{Flux density (Jy) } \\
UT Date	&  2.0\,mm & $\sigma_{2.0}$ &		1.35\,mm &
$\sigma_{1.35}$ &	       0.85\,mm & $\sigma_{0.85}$ &
0.75\,mm & $\sigma_{0.75}$ &    0.45\,mm & $\sigma_{0.45}$ \\ 
{\bf Core}  & & & & &  & & & & & \\
19980214 & & & & &   0.49 & 0.05 & & &  0.23 & 0.04 \\

19980216 & 0.72 & 0.11 & 0.78 & 0.15 & & & & & &  \\

19971002 & & & & & & & 0.51 & 0.05 & \\
 
19970918 & 1.11 & 0.1 & 0.65 & 0.10 & 0.46 & 0.06 & & & & \\
 
19970502 & & & & & 0.53 & 0.05 & & &  0.34 & 0.06 \\ 
{\bf NW Hotspot}   & & & & &  & & & & & \\
19980216 & 2.49 & 0.37 & 1.20 & 0.23 \\
19971002 & & & & & & & 0.67 & 0.06 & \\
19970918 & 2.07 & 0.20 & 0.95 & 0.14 & 0.57 & 0.08 & & & & \\
19970502 & & & & & 0.68 & 0.06 & & & 0.23 & 0.06 \\
{\bf SE Hotspot}  & & & & &  & & & & & \\
19980216 & 3.54 & 0.53 & 1.76 & 0.33 & & & & & & \\
 
19971002 & & & & & & & 0.81 & 0.08 & & \\
 
19970918 & 2.65 & 0.2 & 1.56 & 0.23 & 1.01 & 0.14 & & & & \\
 
19970502 & & & & &  0.95 & 0.05 & & & 0.43 & 0.07

\end{tabular}
\end{minipage}
\end{table*}

In addition to SCUBA observations, $IRAS$ data were reprocessed using HIRES and SCANPI routines at NASA's
Infrared Processing and Analysis Center (IPAC).  The re-reduced 12,
25 and $60\mu{\rm m}$ data
agree well with previous determinations (Fullmer \& Londsdale 1989, Knapp et al.\ 1990,
Golombek, Miley, Neugebauer 1988).  There is some disagreement about the 
100\,$\mu{\rm m}$ value; Golombek et al.\ (1988) indicated that the
upper limit flux density was 1.8\,Jy, however Fullmer \& Lonsdale 
(1989) gave significantly different
upper limit of 8.3\,Jy.  The reason for disagreement is because the published upper limits have been determined in
different ways.  Cygnus-A has a low galactic latitude
($b=6^{\circ}$), and the $100\mu{\rm m}$ flux is contaminated by cirrus
(Low et al.\ 1984).  From the HIRES images it is clear that the cirrus
contamination is dominant and the true
Cygnus A flux at 100\,$\mu{\rm m}$ cannot be retrieved from $IRAS$ data
even with
unusually high iterations of HIRES processing.  Moreover, photometry on the
HIRES images produced a $3 \sigma$ upper limit of 5.1\,Jy. 

\section{Results and Discussion}

The 850, 750 and 450\,$\mu{\rm m}$ SCUBA images (Fig. 1) clearly show the 
two hotspots and the central core of the galaxy. The rms 
uncertainties on the maps are 40\,mJy, 50\,mJy and 85\,mJy 
respectively.  No detections were made at 350\,$\mu{\rm m}$ and a 430\,mJy rms was determined from the map (not shown).  

\subsection{The hotspots}

\begin{figure*}
  \includegraphics[width=14cm]{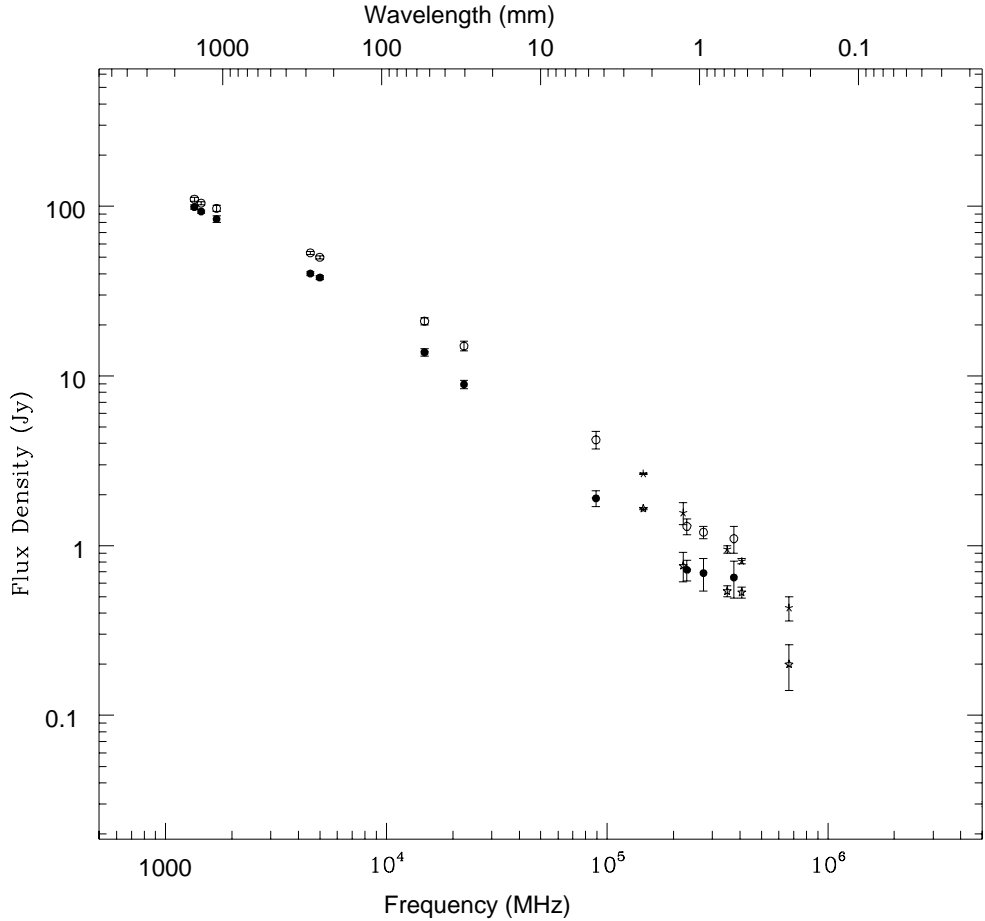}
  \caption{The spectral energy distribution of the hotspots A (open
  circles) and D (solid circles). The 
SCUBA data points are indicated by asterisks and stars.}
\end{figure*}

The spectral energy distribution of the two hotspots 
is shown in Fig. 2. The SCUBA data points are indicated by asterisks (hotspot A) and stars (hotspot D); the other data points (open and solid circles for hotspots A and D, respectively) are taken from Carilli et al.\ (1991). 
The points at 375\,GHz (800\,$\mu{\rm m}$) are from Eales et al.\ (1989) and are also from the JCMT using the previous 
single-pixel continuum instrument, UKT14. 
 
The well observed radio synchrotron spectrum is seen to 
continue smoothly into the submillimetre region with no apparent 
spectral break. The spectral indexes, $\alpha$,  determined from 146\,GHz to
667 GHz are: $\alpha = -1.04 \pm 0.01 $ for hotspot A (NW), and $a 
= -0.99 \pm 0.01 $ for hot-spot D (SE).  In all cases $S_{\nu} \propto
{\nu}^{\alpha}$. There is  
no significant evidence for further spectral steepening due to a high
energy cut-off in the electron distribution.  The excellent
determination of an index of close to $-1.0$ fits extremely well with
the continuous injection of relativistic electrons in the hotspots.  The spectral index of the injected electrons of $-0.5$ gives
support to the origin of the high energy electrons in a shock (Bell 1978). 

If we take an equipartition value for the magnetic field in the hotspots 
of 30nT (Wright \& Birkinshaw 1984), the electrons radiating at 800\,GHz have an energy 
of around $2 \times 10^{10}$\, eV and a synchrotron radiative lifetime of 
$6.6 \times 10^3$\, years.  For a single central source of the electron acceleration in each 
hotspot and an electron diffusion rate of the speed of light, the 
electrons responsible for the 800\,GHz synchrotron radiation can, 
at most, travel a radial distance of 2\,kpc. Interferometer 
measurements at 89\,GHz (Wright \& Birkinshaw 1984) reveal that the hotspots are 
spatially resolved and are between 2 and 3\,kpc in size. Therefore 
if the hotspots are the same size at 800\,GHz, there is barely adequate 
time for the highest energy electrons to fill the volume even with a
diffusion speed of order c. If the speed is substantially less than c, 
then multiple acceleration sources (shocks) are required in order 
to explain the observations. To test this hypothesis further 
requires better determination of the magnetic field and 
interferometer measurements at 800\,GHz. These will be possible 
with the introduction of the Submillimeter Wavelength Array (Moran 1996) on Mauna Kea around 
2000.

\subsection{The core component}

A power-law spectral index of $\alpha = -0.6 \pm 0.1$ has been determined
between 146 and 677\,GHz for the core.  This is much steeper than $\alpha
\approx -0.1$ determined by Wright \& Birkinshaw (1984) and Salter et
al. (1989) respectively between 10 and 89\,GHz and 89 and 230\,GHz.
Indeed, given the quality of the data it is not clear whether a
power law (rather than a smoothly
curving function) is, in fact, the best fit to the data.  On the other hand, in the unification scenario we
suspect the radio-core of Cygnus-A to be a severely misaligned jet and
therefore the spectrum to be typical of an unboosted blazar jet spectrum.  These have
been observed in the millimetre to submillimetre (Gear et al.\
1994), so we assume that a power-law is the simplest interpretation.
In this case there is tentative evidence for some steepening above
600\,GHz, but the data are inadequate  to determine whether this is
the 0.5 break of a continuous injection model, or, whether it is the
onset of depletion of high energy electrons.  Further observations are
underway to answer this question.

The $IRAS$ measurements clearly suggest 
thermal emission from dust in the central galaxy, although the 
temperature and hence mass of the emitting dust are very 
uncertain from the $IRAS$ values alone.  The new submillimetre 
data points constrain the non-thermal contribution to the $IRAS$ (and
the much awaited $ISO$) fluxes.  In particular the 667\,GHz
measurement, constrains the temperature of emitting dust to within a
factor of two.  Preliminary $ISO$ data (Polletta, private
communication) at $170\,{\mu \rm{m}}$ suggests that the dust
temperature may be $\approx 50$\,K (which is well within our
constraints -- see below).  However, this data
point still has a large calibration uncertainty at the current time
and is not included in the present discussion.

\subsection{Cold dust in Cygnus-A?}

\begin{figure*}
  \includegraphics[width=14cm]{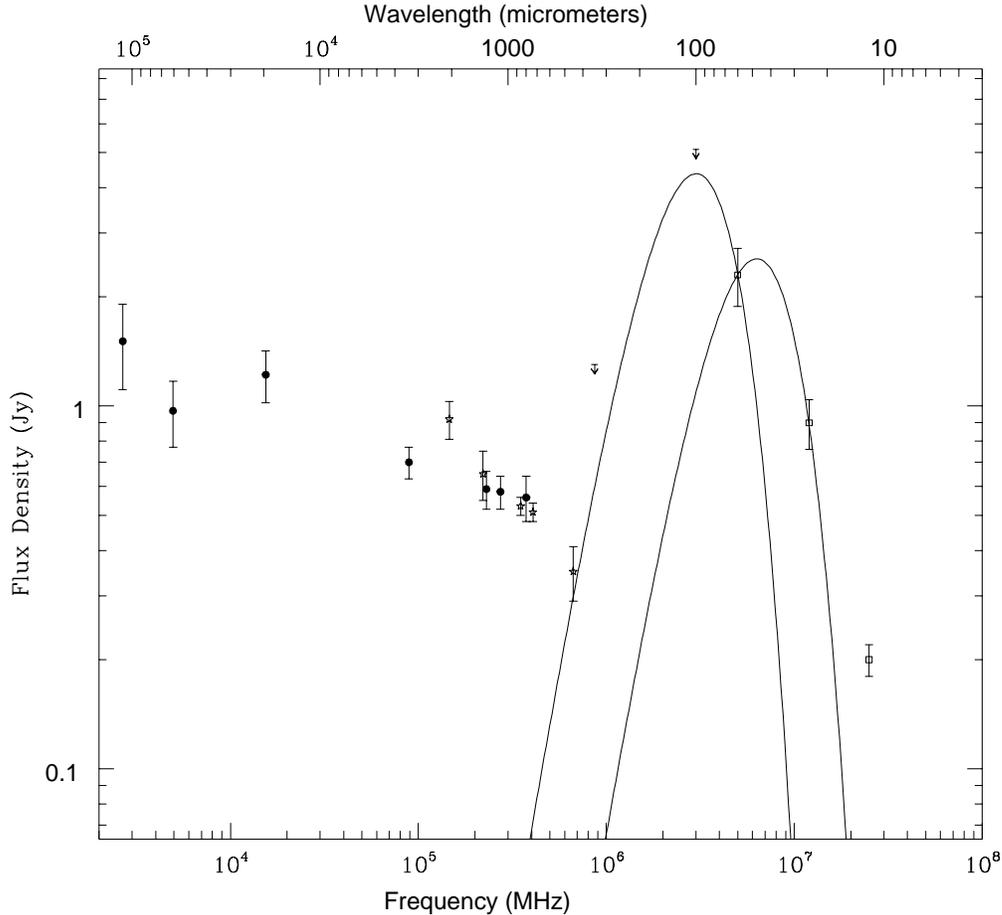}
  \caption{The spectral energy distribution of the central core. The 
SCUBA data points are indicated by open stars. The upper limits are
  both 3$\sigma$; the 350\,$\mu$m is from SCUBA and the 100\,$\mu$m is from
  reprocessed $IRAS$ data.  The two 
curves represent emission from greybodies with temperatures of 
37K and 85K with an emissivity index, $\beta$, of 1.3. }
\end{figure*}

Assuming a reasonable value for the dust emissivity index, $\beta$, of 
1.3, the maximum temperature that appears to fit the $IRAS$ data 
is 85K, while the lower temperature is constrained by the 
SCUBA measurement at 450 microns to 37K. Fig. 3 shows both 
these temperatures. Changing $\beta$ between 2.0 and 1.0 makes 
very little difference to the derived temperatures.  The mass of 
emitting dust $M_{\rmn{d}}$ can then be derived from a simple model 
adapted from Hildebrand (1983), where  

\begin{equation}
M_{\rm d}= \frac{{S_{\nu} D^2}}{{k_{\rm d} B(\nu ,T)}};
\end{equation}
and $S_{\nu}$ is the measured flux, $D$ the distance to the source,
$B({\nu} ,T)$ the Planck function and $k_{\rm d} = 3Q_{\nu}/4a\rho$ the
grain mass absorption  coefficient where $a$ and $\rho$ are respectively
given by the grain radius and density.  In the far-infrared and
submillimetre, $k_{\rm d}$ is not well defined since it depends primarily on
the uncertainty of the grain emission efficiency, $Q_v$ (Draine 1990).
We assume values of
$k_{\rm d}^{450\,\mu {\rm m}} = 0.25\,{\rm m}^2 {\rm kg}^{-1}$ and
$k_{\rm d}^{60\,\mu {\rm m}} = 3.3\,{\rm m}^2 {\rm kg}^{-1}$, which yield dust masses of $1.0 \times 10^8 M_{\odot}$ for ${\rm T}=37$\,K 
and $1.4 \times 10^6 M_{\odot}$ for ${\rm T}=85$\,K. 
    
Hughes et al.\ (1994) estimated the dust mass within the inner 
1kpc of the nucleus of the starburst galaxy M82 to be between $1.8 \times 10^{6} 
M_{\odot}$ and $3.6 \times 10^{6} 
M_{\odot}$ for T$=48$K \& T$=30$K respectively.  The minimum estimated 85\,K dust mass in 
Cygnus-A is remarkably close to this, and, though only a tantalising hint, suggests Cygnus-A
core properties are similar to those of classical nuclear starburst
galaxies.  However, our upper limit to the dust mass for Cygnus-A is considerably higher than the dust mass 
in M82, even when considering the mass contribution seen in the 
low brightness flux outside the inner 1kpc of the nucleus, which 
is measured to be no more than 10\% of the nuclear mass (Leeuw et
al.\ in preparation, Hughes et al.\ 1994).

From our deep images of the extended dust emission in
M82 (Leeuw et al. in preparation), we can safely say that if we move M82 to the distance
of Cygnus-A, the emission is unresolved in the SCUBA beams.  Our recently
obtained deep SCUBA images of the star forming galaxy Arp220 (at 76Mpc) are
also unresolved.  

Our upper limit for the dust mass in Cygnus-A is comparable to the dust masses for Arp220 and
another star forming galaxy NGC6240, which are 4.5$\times 10^{7} 
M_{\odot}$  and $3.0 \times 10^{7} 
M_{\odot}$ respectively (Klaas et al.\, 1997).  The comparison is
interesting because, from unusually high hot molecular gas emission ($1.0 \times 10^{34}\,{\rm W}$), Ward et al.\ (1991) likened the core of Cygnus-A to Arp220
and NGC6240.  At the
same time, Ward et al.\ (1991) 
noted that the implied mass of the hot molecular gas led to mass ratio
upper limit of hot-cold ${\rm H}_2$ that was not unusual in AGN.   We note that
while the lower and upper dust mass limits for Cygnus-A are,
respectively, comparable to the nuclear starburst
galaxies M82 and the extranuclear starburst galaxies Arp220 and
NGC6240, the upper dust mass limit is also similar to the 
masses for the Radio Quiet Quasars IZw1, Mrk1014 and Mrk376, which are 7.6$\times 10^{7} 
M_{\odot}$, $6.3\times 10^{8} 
M_{\odot}$ and $5.8\times 10^{7} 
M_{\odot}$ respectively (Hughes et al.\
1993).   SCUBA observations at $350\,{\mu m}$ will provide better
constraint on this  upper limit and perhaps clarify the role of dust
in models of the Cygnus-A core.

\section{Summary}

We have made detailed photometric and imaging observations of 
the two hotspots and the radio core of Cygnus-A. The spectral 
index of the hotspots extends smoothly to about 677\,GHz with no 
evidence of spectral steepening. If the diffusion speed for the 
electrons responsible for this emission is about c, then a single particle 
acceleration mechanism can be responsible for the hotspot 
synchrotron emission. On the other hand, if the diffusion speed is 
significantly less than c, then multiple acceleration locations are 
probably required unless the magnetic field strength is 
significantly less than the equipartition value. Future observations 
using submillimetre interferometers should narrow-down the 
parameter space in terms of the size of the emitting region.

The radio core has a much flatter spectral index, more indicative 
of a relativistic jet, and as such could be the standing shock at the 
entrance of the parsec-scale jet. There is some evidence for spectral 
steepening, perhaps suggesting that at these energies, the lifetime of the 
radiating particles is sufficiently short that there is inadequate 
replenishment available. There is no evidence for thermal 
emission from dust at the highest frequency of 667\,GHz and this, 
along with re-reduced $IRAS$ data restrict the temperature of the 
emitting dust to between 37K and 85K, with corresponding dust 
masses of $1.0 \times 10^8 M_{\odot}$ 
and $1.4 \times 10^6 M_{\odot}$. The lower dust mass limit is
comparable to the nuclear starburst galaxy M82 and the upper limit
to the extranuclear starburst galaxies Arp220 and NGC6240 as well as the Radio
Quiet Quasars IZw1, Mrk1014 and Mrk376. 

\section*{Acknowledgments}

LLL thanks the University of Central Lancashire for
a full-time research studentship.  The JCMT is operated by the Joint Astronomy Centre, on behalf of the
UK Particle Physics and Astronomy Research Council, the Netherlands
Organization for Scientific Research and the Canadian National
Research Council.  This research has made use of HIRES and SCANPI
routines at NASA/IPAC which is operated by the Jet Propulsion
Laboratory, Caltech, under contract to the National Aeronautics and
Space Administration.

\label{lastpage}

\end{document}